# The Rigidity Dependence of the Diffusion Coefficient in the Heliosheath and an Explanation of the Extreme Solar Modulation Effects for Cosmic Ray Electrons from 3-60 MeV Observed at Voyager 1

William R. Webber[1], Thomas E. Harrison[1], Nand Lal[2], Bryant Heikkila[2] and Tina L.Villa[1]

1. New Mexico State University, Astronomy Department, Las Cruces, NM 88003, USA
2. NASA/Goddard Space Flight Center, Greenbelt, MD 20771, USA

# ABSTRACT

We believe that the extreme solar modulation of 3-60 MeV Galactic electrons measured by Voyager in the heliosheath and the interpretation of this new data in terms of the rigidity dependence of the diffusion coefficient at low rigidities presented in this paper represents a major step in understanding diffusion theory as it applies to energetic particles. This description uses electron spectra measured at 5 different epochs and distances within the heliosheath. The diffusion dependence needed to explain the remarkable solar modulation effects observed for both electrons and higher rigidity protons as one progresses from the heliopause inward by ~25 AU to the termination shock really has two distinct rigidity regimes. Above a rigidity ~$P_c$ the diffusion coefficient has a dependence ~$\beta P$, the modulation is ~$P$ and its magnitude increases linearly with radius in AU according to the integral of $\frac{dr}{K}$. This integral defines a potential, $\phi$, called the modulation potential, thus explaining the proton variations. At rigidities <$P_c$, the diffusion coefficient is ~$\beta$ and independent of rigidity. The modulation is also independent of rigidity but its magnitude depends on the modulation potential, thus explaining the electron modulation. One needs both electron and proton observations, together, to recognize the physical description of the solar modulation process. For the first time we have been able, using proton data at high rigidities and electron data at low rigidities, to put together a picture of the high and low rigidity diffusion coefficients and how they affect energetic particles in an astrophysical scale environment.

# 1. Introduction

The outermost region of the heliosphere, the heliosheath, is an ongoing mystery in many ways. This region is a transition region between the heliospheric termination shock (HTS) at which point the pressure of the outward moving solar wind plasma and magnetic fields equals that of interstellar fields and plasma, and an outer boundary, the heliopause, where solar dominated effects rather suddenly give way to galactic dominated field and plasma effects. Voyager 1 and 2 measurements have placed the HTS distance from the Sun at ~85-95 AU with this distance depending on the 11 year solar activity cycle (Stone, et al., 2008). V1 crossed the HP at a distance ~121.5 AU thus determining a heliosheath thickness ~26.5 AU (Cummings, et al., 2016). In the heliosheath the solar magnetic field becomes larger and more variable, the plasma much more turbulent than inside the HTS, with outward moving solar originating shocks and an enormous acceleration of nuclei and electrons from KeV to tens of MeV in energy.

This region also has a profound effect on the intensity of galactic cosmic rays entering the heliosphere and reaching the Earth, described as solar modulation (Webber, Stone & Cummings, 2017). In many of the previous studies of this modulation these outer heliosphere effects were not even considered separately in the solar modulation calculations, one goal of which is to determine the intensity and spectra of galactic cosmic rays outside the HP using measurements near the Earth, a step in the process of understanding the enigmatic origin and acceleration of cosmic rays.

Voyager 1 has continued on for over 6 years beyond the HP corresponding to ~20 AU of outward travel during which time it has measured in some detail these galactic cosmic ray nuclei of all charges H through Fe and not the least, electrons from MeV to GeV energies in some cases. From the two passes of V1 and V2 through the heliosheath starting in 2005 and 2007 we have learned that this region is an important contributor to solar modulation (Webber, Stone and Cummings, et al., 2017). At the 250 MeV energy for protons observed by Voyager 1, the proton intensity increased by a factor ~4 in the heliosheath between the HTS and HP. For 15 MeV electrons this factor is ~100! In effect virtually all of these low energy electrons are excluded from the region inside the HTS as a result of solar modulation effects in the heliosheath (Webber, Lal & Heikkila, 2018).

For protons and He nuclei we have found that the radial intensity profiles the heliosheath (and also throughout the entire heliosphere) could be described in terms of a modulation potential φ, in MV (Webber, Stone & Cummings, et al., 2017), which is very similar to that produced by an electric field. The description of this simplification of the complete transport equations, and its range of validity, has been given by Gleeson & Axford, 1968. This simple description arises from the fact that the overall spherically symmetric solar modulation in the heliosphere appears to follow the description provided by Liouville's theorem relating to the constancy of the particle density and momentum in phase space. Of course there are significant deviations from this simple picture due to structural features in the heliosphere such as the tilt of the heliospheric current sheet, and also for the solar polarity changes which induce a 22 year cycle in the solar modulation process, but these other processes do not appear to dominate at the higher energies.

In fact, the same value of the modulation potential, φ, obtained from the Voyager studies throughout the heliosphere at rigidities ~1 GV also gives a good description of the historical neutron monitor observations of cosmic ray modulation effects at several GV at the Earth that have been carried out over the last 70 years (Webber, Stone & Cummings, et al., 2017; Usoskin, et al., 2015).

For electrons we now have detailed intensity measurements during the same heliosheath pass of V1, and covering the energy range 3-60 MeV. These measurements have been described in Webber, Lal & Heikkila, 2018.

Our goal in this paper is to interpret these new and unique measurements of electrons following the same approach in terms of the modulation potential, φ, used by Webber, Stone & Cummings, et al., 2017, to interpret protons and helium nuclei observations at a higher energy made by V1 in the heliosheath.

## 2. The Electron Observations and Their Interpretation

Figure 1 shows the electron spectra between 3-60 MeV measured at V1 at 5 different times during its passage through the heliosheath. At each of these times the proton intensity is also measured and the corresponding modulation potential, φ, for protons and He nuclei has been determined. These values for the modulation potential are 60, 80, 100, 140 and 200 MV for 5 time periods starting just inside the HP and ending just outside the HTS (Webber, Stone, Cummings, et al., 2017).

In Figure 1 we also show the calculated electron intensities that would be expected at lower energies if the dependence of the diffusion coefficient, K was ~βP, the same as is applicable at the higher energies measured by protons and Helium nuclei and continued down to lower energies (where P = rigidity = E for electrons). These calculations are based on the Gleeson & Axford, 1968, formulation of the modulation for electrons

$$\left(\frac{dj}{dE}\right)_1 = \left(\frac{dj}{dE}\right)_2 (E + \phi) \left[\frac{E}{E + \phi}\right]$$

where $\left(\frac{dj}{dE}\right)_2 (E + \phi)$ is the interstellar electron intensity as measured by Voyager (Cummings, et al., 2016) and extended to higher energies (Webber & Villa, 2017), and $\left(\frac{dj}{dE}\right)_1$ is the electron intensity measured in the heliosphere and $\left[\frac{E}{E + \phi}\right]$ is the Liouville factor.

Comparing these simple calculations using the βP dependence of K extended to lower rigidities and the measurements in Figure 1, we see that the calculations lie well below the measurements for low levels of solar modulation, e.g., $\phi$ = 60, 84, 100 MV, as measured for protons. But at higher levels of modulation, $\phi$ = 140 and 200 MV for protons, the calculations agree with the measurements at the higher energies, before plunging to much more extreme levels of modulation at ~10 MeV and below.

In the original formulation for a spherically symmetric solar modulation, Gleeson & Axford, 1968, presented the following relationship

$$\Phi = \frac{ZeP}{K_2\ (P)}\ \phi = \frac{\alpha E}{3} \int \frac{V\ (r,t)}{K_1\ (r,t)}\ dr$$

where $\alpha = \phi/P_c$ and V = solar wind speed . They found that when $K_1$ ~βP, there was a particularly simple solution where the modulation potential $\phi$=const. at all rigidities > $P_c$. In this case, ϕ is described as the force field potential. This was viewed as applying mainly above a $P_c$ ~1 GV.

Lezniak & Webber, 1971, (see also Lezniak, 1969) examined the solutions to the above relationship where K is now ~β as applying to the diffusion at lower rigidities. In this case the potential ϕ is now a function of rigidity as follows: when $P > P_0$

$$\Phi = Ze\ \phi = \text{const. (Gleeson \& Axford force field solution)}$$

and when P is < ($P_0$-α)

$$\Phi = P_0\ \cosh\ (P_0 + \alpha)$$

The potential, ϕ, as a function of rigidity, in this case, is shown in Figure 2 for the 5 values of the potential derived from the proton data at higher rigidities. It is seen that ϕ rapidly decreases below $P_0$ with a dependence which is ~P. So the low energy modulation will be less than expected for a constant value of ϕ at higher rigidities.

In Figure 3 we show the total electron modulation actually measured at V1 in the heliosheath (Webber, Lal & Heikkila, 2018), expressed as $\ell n$ ($J_{LIS}$/J). At higher rigidities the modulation for a given ϕ is determined by the βP dependence of the diffusion coefficient and is ~P. At lower rigidities, below a value of P = $P_c$, the value of $\ell n$ ($J_{LIS}$/J) appears to become almost constant below a lower rigidity for each specific value of ϕ determined above ~1 GV. Also in Figure 3 we show a calculation of the expected modulation which uses the Lezniak & Webber, 1971, formulation, described by the equations above, for the rigidity dependence of ϕ below $P_c$. These calculations are made for each value of ϕ as determined from protons, which in turn leads to specific values of $P_c$ at the different times as shown in Figure 2. The fit to the measured values of the electron modulation at Voyager, $\ell n \left( \frac{J_{LIS}}{J_e} \right)$, between rigidities of 10-60 MV for each time interval is good.

At rigidities below 10 MV there appears to be a turn up in the actual measured electron spectra in the heliosheath, probably unrelated to the solar modulation, but to a new unrecognized component of low rigidity electrons having its origin in the heliosheath (Webber, Lal & Heikkila, 2018).

And finally in Figure 4, we show the diffusion coefficient that is required to fit the data in the heliosphere at each of the 5 times when the modulating potential for protons has been measured, moving from just inside the HP, where the value of $\phi$ from the proton modulation is 60 MV, to a point where $\phi$ has increased to ~200 MV, just outside the HTS. This figure illustrates the fact that the diffusion coefficient below each value of $P_c$ remains almost a constant, independent of rigidity, as the values of $\phi$ change from 60 MV to 200 MV. Furthermore the value of the diffusion coefficient above $P_c$ increases in a linear fashion with regard to the distance from the heliopause. This is consistent with the description of $\phi$ in terms of the integral $= \int \frac{V dr}{K_1(r,P)}$ in the Gleeson & Axford, 1968, relationship between $\phi$ and K(r) and when V (r) = const (force field model).

This behavior of the modulation potential between the HTS and the HP is also consistent with the variation in potential between two charged spheres at different potentials which are $V_1$ at the HTS and $V_2$ at the HP with radaii $r_1$ and $r_2$.

The break in the diffusion coefficient by a power ~P at $P_c$ as shown in Figure 4, and the constancy of the value of the diffusion coefficient below the rigidity $P_c$, which has a value in the range between 50-165 MV, are important factors for the propagation of energetic particles, both electrons and nuclei, not only in the heliosphere as illustrated above but, appropriately scaled, in all astrophysical environments, large or small, including the interstellar medium.

### 3. Summary and Conclusions

During the passage of V1 through the heliosheath, considerable intensity changes related to solar modulation were observed. For protons with energies ~250 MeV, an intensity increase of a factor ~4 was observed between the inner limit of the heliosheath, the HTS, and the outer limit, the HP (Webber, Stone & Cummings, 2017). For electrons from 3-60 MeV this increase was as much as a factor 100 (Webber, Lal & Heikkila, 2018).

We have explained these intensity increases using a variant of the Gleeson & Axford, 1968, description of solar modulation in terms of a modulation potential, $\phi$, in MV. This potential acts much like an electric field potential at higher rigidities. This potential results from a solution to the 3-D transport equation by Gleeson & Axford, 1968 with limits based on experimental observations and is equivalent to the application of Liouville's theorem relating to the constancy of the intensity and momentum spectra of particles at two locations in space (outside and inside the heliosphere, for example) in the calculation of solar modulation.

We find that the intensity changes of protons and electrons that are observed can both be simultaneously explained by using a diffusion coefficient that is $\sim \beta P$ above a rigidity $P_c$ and $\sim \beta$ below $P_c$. Above $P_c$ there is a single value of the modulation potential, $\phi$, that is independent of P and leads to a modulation that is $\sim P$. Below $P_c$ where $K \sim \beta$, the diffusion coefficient itself becomes independent of P and the modulation itself, $\ell n \left( \frac{J_{LIS}}{J_e} \right)$, also becomes independent of P but now with a much larger value which depends on the modulation potential.

For values of the modulation potential which range from 60 MV to 200 MV, we obtain values of $P_c$ between 165 to 50 MV respectively, below which the diffusion coefficient is independent of P. The quantity $\phi \cdot P_c$ is therefore almost a constant.

The observations and calculations reported here represent the first observation in an astrophysical plasma of a pattern of diffusion coefficient changes at low rigidities related to the "pile up" of the turbulence cascade which leads to rigidity dependent diffusion of particles at higher rigidities (e.g., Ptuskin, et al., 2006).

This observation represents only one pass through the heliosheath. On the V2 pass the modulation potential just outside the HTS, as determined from protons, was only 150 MV instead of 200 MV as observed at V1. So the total magnitude of the heliosheath modulation is probably solar cycle dependent, but its physical description may be similar.

**Acknowledgements:** The authors are grateful to the Voyager team that designed and built the CRS experiment with the hope that one day it would measure the galactic spectra of nuclei and electrons. This includes the present team with Ed Stone as PI, Alan Cummings, Nand Lal and Bryant Heikkila, and to others who are no longer members of the team, F.B. McDonald and R.E. Vogt. Their prescience will not be forgotten. This work has been supported throughout the more than 40 years since the launch of the Voyagers by the JPL.

# FIGURE CAPTIONS

**Figure 1:** Galactic electron intensities from 3 to 60 MeV measured in the LIM and also at 5 other times and radial locations in the heliosheath by V1. The calculated electron intensities at these locations that are shown here are determined using the method described in the text where $\phi$ the modulation potential is obtained from the proton intensity modulation measured at each location.

**Figure 2:** The values of the modulation potential, $\phi$, as a function of rigidity where $K = \beta P$ above $P_c$ and $K = \beta$ below $P_c$ for the 5 levels of solar modulation observed in the heliosheath.

**Figure 3:** Total measured electron modulation, $\ell n \left( \frac{j_{LIS}}{j} \right)$ (red dots), as a function of rigidity for the 5 levels of modulation observed in the heliosheath. The calculated modulation for the 5 levels using the Lezniak & Webber, 1971, formulation for the variation of $\phi$ below $P_c$ is shown.

**Figure 4:** The values of the diffusion coefficient at the times when the modulation potential at higher rigidities is determined to be 60, 100, 140 and 200 MV from the proton modulation. The red dots correspond to the values of $P_c = 165, 95, 70$ and 50 MV respectively for these time intervals. The numbers between the separate lines above $P = P_c$ are the radial distances in AU between the time intervals.

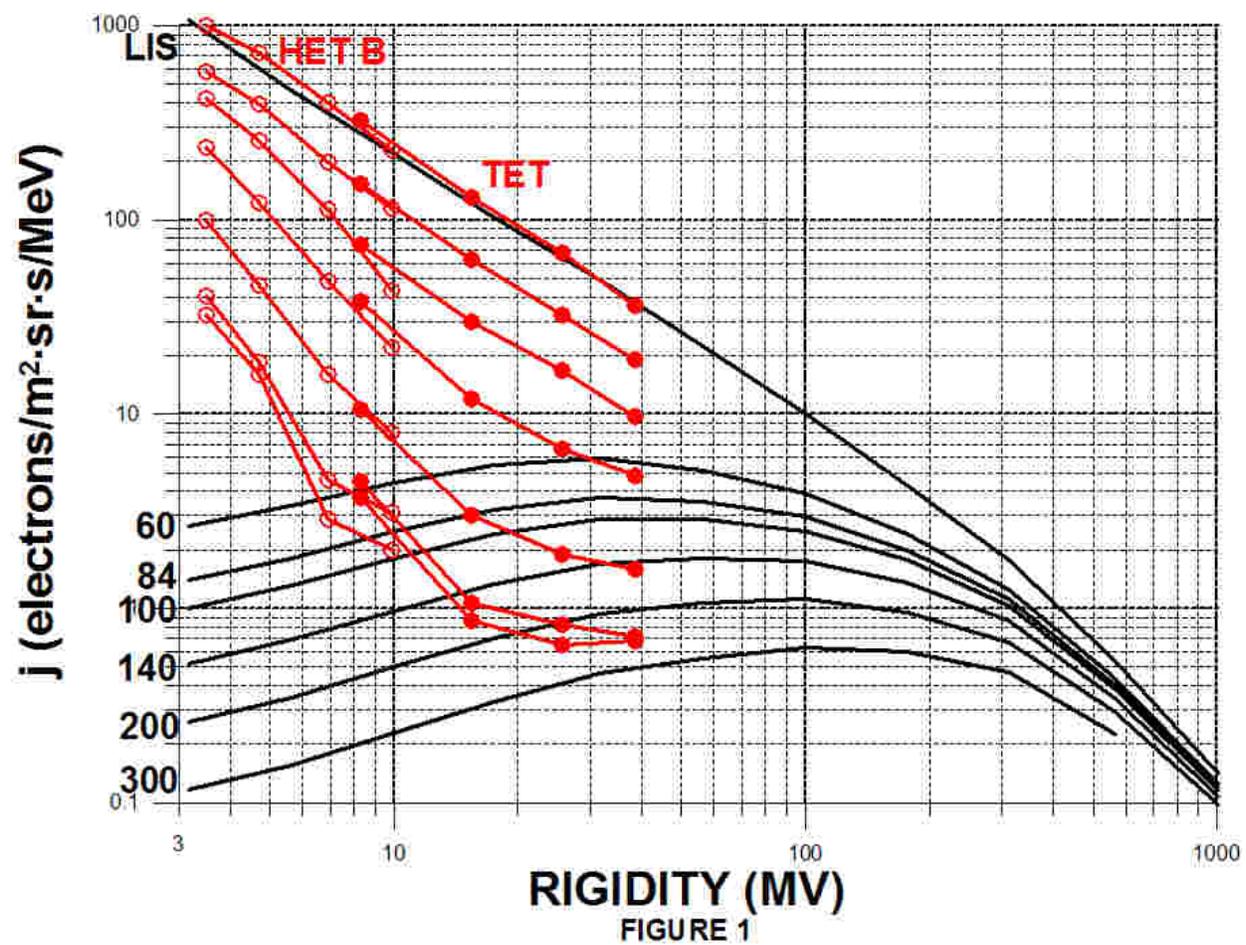

FIGURE 1

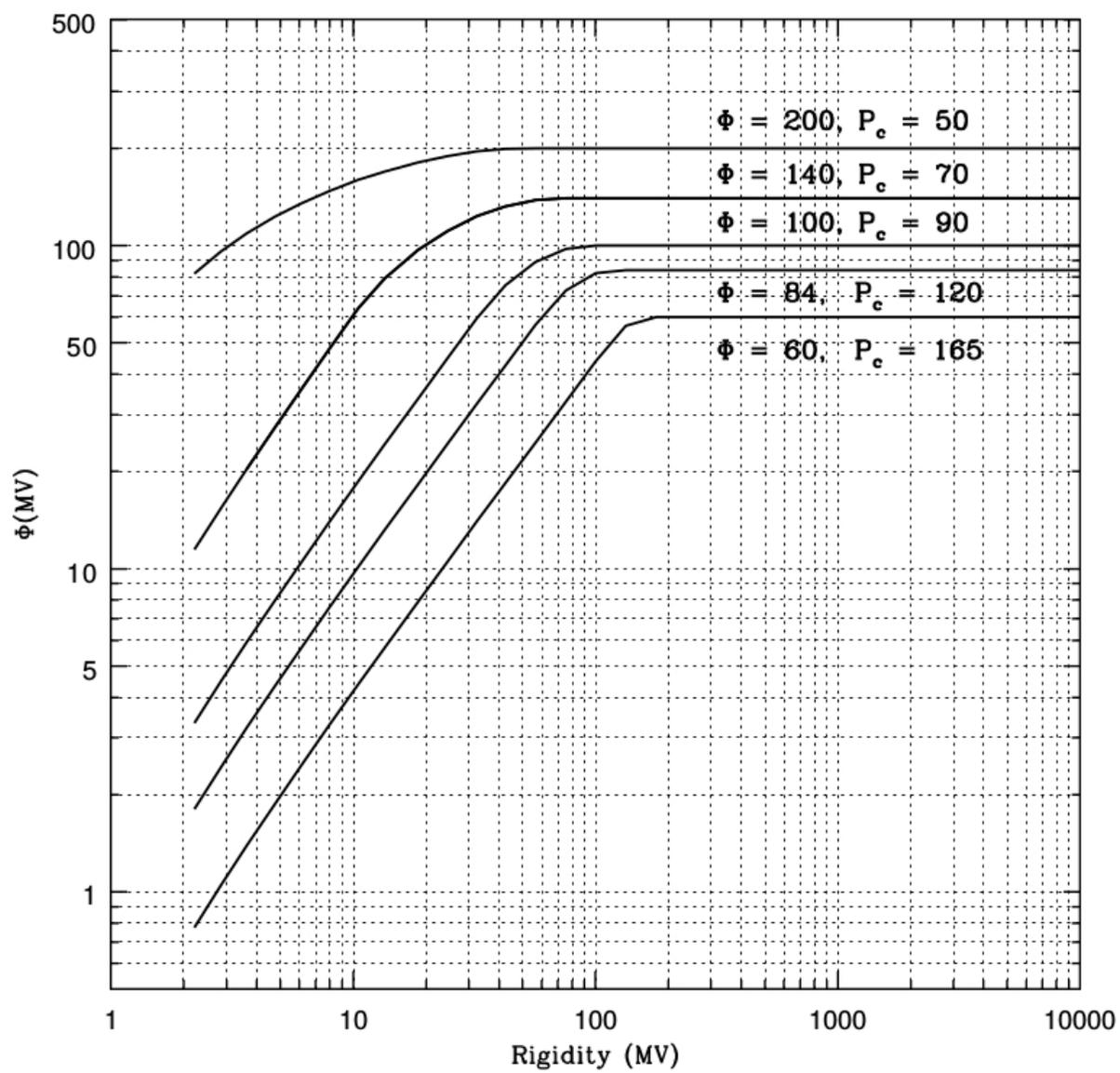

FIGURE 2

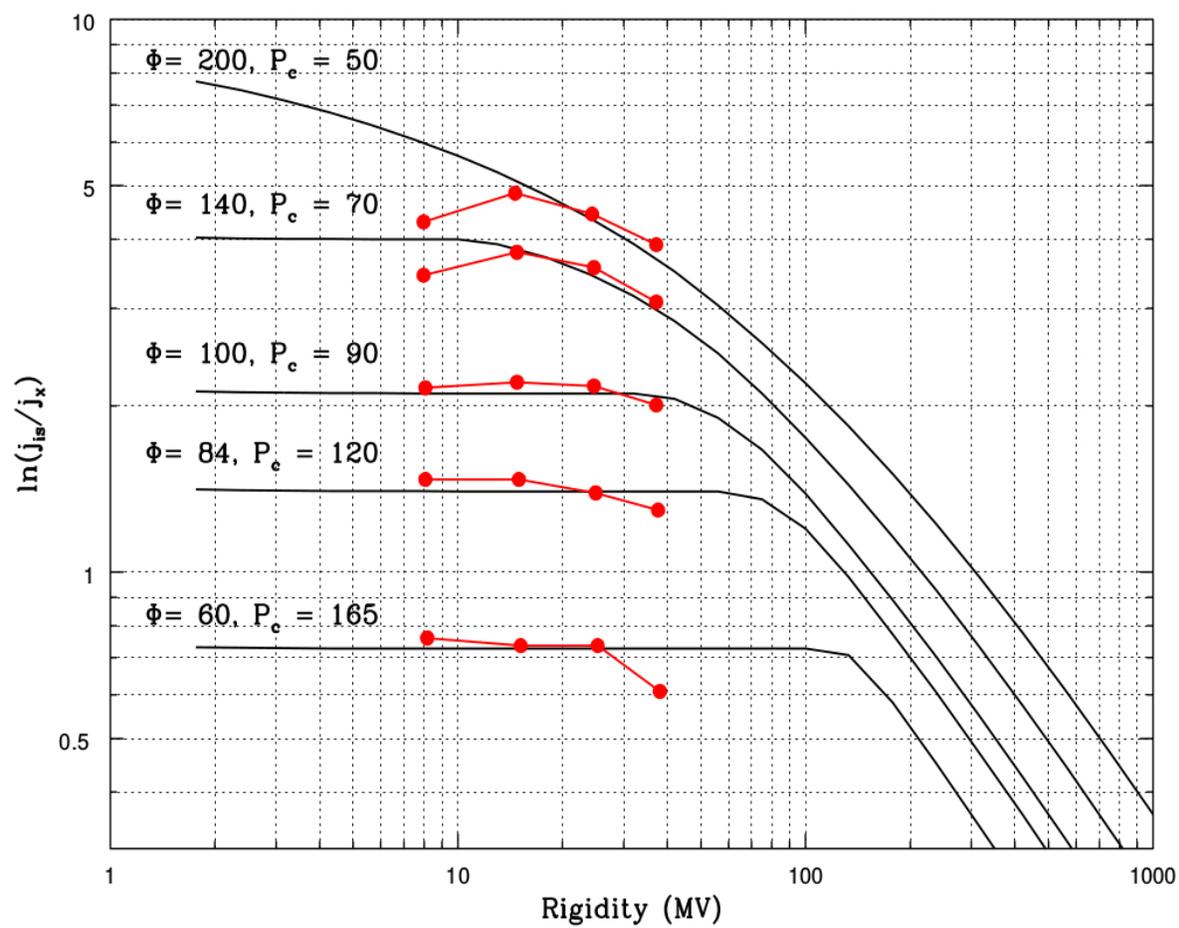

FIGURE 3

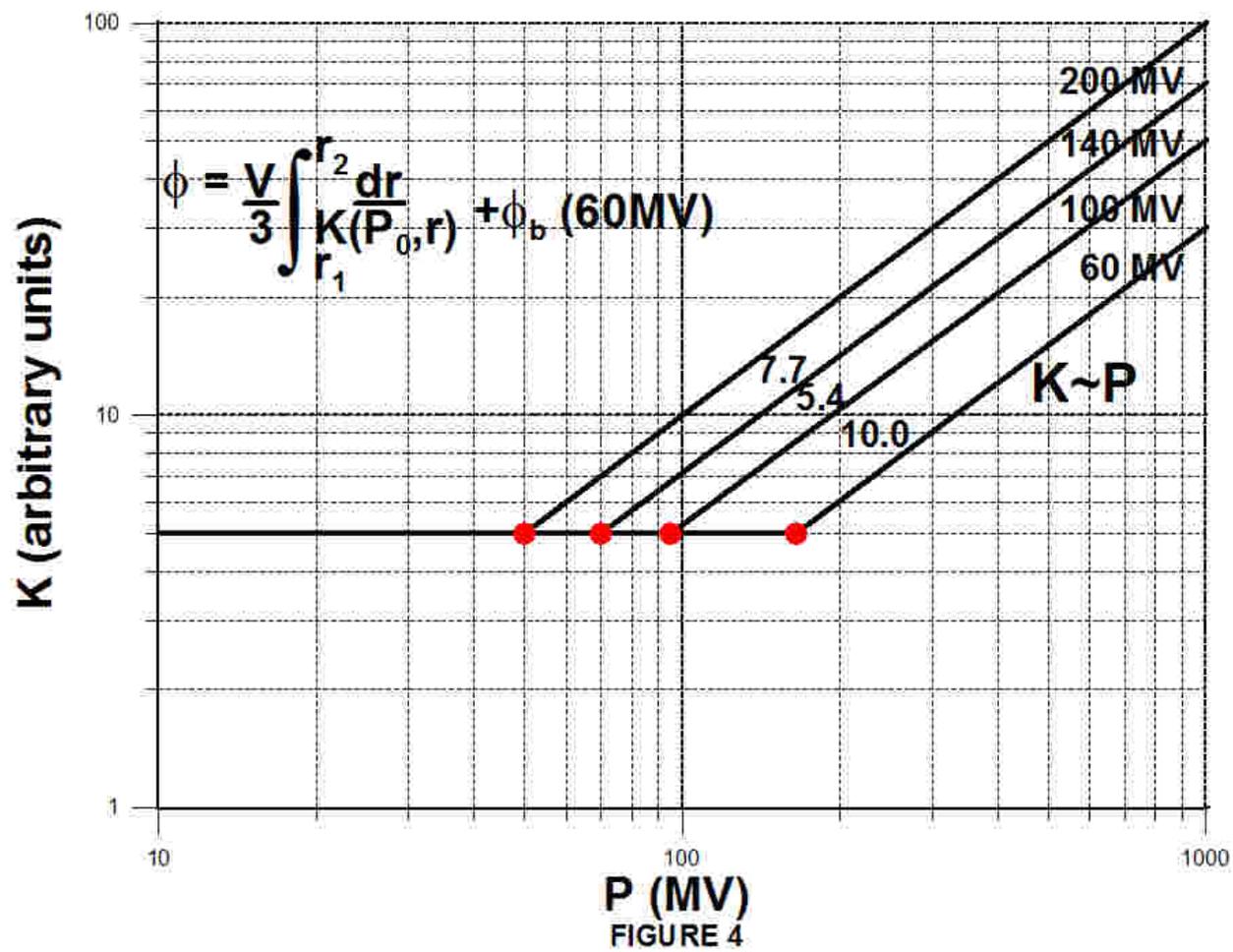

FIGURE 4